\documentclass[fleqn,twoside]{article}
\usepackage{espcrc2}
\usepackage{graphicx}

\newcommand{\AmS}{{\protect\the\textfont2
  A\kern-.1667em\lower.5ex\hbox{M}\kern-.125emS}}

\title{Radiatively Induced Neutrino Mass Matrix \\
in a SUSY GUT Model with $R$-Parity Violation}

\author{Yoshio Koide\address[US]{Department of Physics, 
        University of Shizuoka, 
        52-1 Yada, Shizuoka, Japan 422-8526}%
        \thanks{E-mail address: koide@u-shizuoka-ken.ac.jp},
         and
        Ambar Ghosal\addressmark[US]
        \thanks{ E-mail
        address : ambar@anp.saha.ernet.in}
        \thanks{Present address:
         Saha Institute of Nuclear Physics,
         1/AF Bidhannagar, Kolkata 700 064, India.}}

\begin{document}

\begin{abstract}
In order  to evade the proton decay which appears when
the Zee model is embedded into a SUSY GUT scenario with 
$R$-parity violation, a new idea based on a discrete  
Z$_2$ symmetry is proposed.  Under the symmetry Z$_2$,
the quark and
lepton mass matrices are tightly constrained.
The admissible form
of the radiatively-induced neutrino mass matrix
is investigated.
\vspace{1pc}
\end{abstract}

\maketitle




\section{Introduction}

The Zee model \cite{Zee} is one of
promising models of neutrino mass
generation mechanism, because the model has only 3 free 
parameters and it can naturally lead to a large neutrino 
mixing \cite{Zee2}, especially, to a
bimaximal mixing \cite{Jarlskog}.
However,  the original Zee model is not on a framework of 
a grand unification theory (GUT), and moreover, 
it is recently pointed out \cite{Koide} that
the predicted value of
$\sin^2 2\theta_{solar}$ must be satisfied the relation 
$\sin^2 2\theta_{solar} > 0.99$ for 
$\Delta m_{solar}^2/\Delta m_{atm}^2 \sim 10^{-2}$, i.e.,
$$
\sin^2 2\theta_{solar} \geq
1 - \frac{1}{16}\left(
\frac{\Delta m^2_{solar}}{\Delta m_{atm}^2}\right)^2 \ .
\eqno(1.1)
$$
The conclusion cannot be loosened 
even if we take the renormalization group equation
(RGE) effects
into consideration.

The simple ways to evade the constraint (1.1) may be 
as follows:
One is to consider \cite{ZeeH2} that the Yukawa vertices of 
the charged leptons can couple to both scalars
$\phi_1$ and $\phi_2$.
Another one \cite{Zee_nuR} is to introduce a single 
right-handed neutrino $\nu_R$ and
a second singlet Zee scalar $S^+$.
Also, a model with a new doubly charged scalar $k^{++}$ is 
interesting because the two loop effects in such a model 
can give non-negligible contributions to the neutrino
masses \cite{k++}. 
As another attractive model, there is an idea \cite{R_SUSY}
that in an $R$-parity violating supersymmetric (SUSY) model
we identify the Zee scalar  $h^+$ as the slepton 
$ \widetilde{e}_R$. 
Then, we can obtain  additional contributions from the 
down-quark loop diagrams to the neutrino
masses, so that such a model
can be free from the constraint (1.1). 

However, these models have not been
embedded into a GUT scenario.
As an extended Zee model based on a
 GUT scenario, there is,
for example, the Haba-Matsuda-Tanimoto
model \cite{Haba}. 
They have regarded the Zee scalar $h^+$ as 
a member of the messenger field $M_{10} +\overline{M_{10}}$ 
of SUSY-breaking on the basis of an SU(5) SUSY GUT.
However, their model cannot escape from the constraint (1.1)
because the radiative masses are only induced by the charged
lepton loop diagrams.

In the present paper, we will investigate an
extended Zee model
which is based on a framework of a SUSY GUT with  
$R$-parity violation, 
and which is free from the severe constraint (1.1). 
Usually, it is accepted that SUSY models with $R$-parity 
violation are incompatible with a GUT scenario, 
because the $R$-parity violating interactions induce 
the proton decay.
In order to suppress the proton decay due to the $R$-parity 
violating terms, we will introduce a discrete Z$_2$ symmetry.

\section{How to evade the proton decay} 

We identify the Zee scalar $h^+$ as the slepton 
$\widetilde{e}_{R}^+$
which is a member of SU(5) 10-plet sfermions 
$\widetilde{\psi}_{10}$. 
Then, the Zee interactions correspond to the following 
$R$-parity violating interactions 
$$
\lambda_{ij}^k(\overline{\psi}_{\overline{5}}^c)_i^A
(\psi_{\overline{5}})_j^B
(\widetilde{\psi}_{10})_{kAB}
$$
$$
=\frac{1}{\sqrt{2}}\lambda_{ij}^{k}\left\{
\varepsilon_{\alpha \beta \gamma}
(\overline{d}_R)_i^{\alpha}(d_R^c)_j^{\beta}
(\widetilde{u}_R^{\dagger})_k^{\gamma} \right\}.
$$
$$
-[(\overline{e}_L^c)_i(\nu_L)_j -
(\overline{\nu}_L^c)_i(e_L)_j]
(\widetilde{e}_R^{\dagger})_k 
$$
$$
- [(\overline{e}_L^c)_i(d_R^c)_j^{\alpha} -
(\overline{d}_R)_i^{\alpha}
(e_L)_j] (\widetilde{u}_L)_{k\alpha} 
$$
$$\left.
+ [(\overline{\nu}_L^c)_i(d_R^c)_j^{\alpha} -
(\overline{d}_R)_i^{\alpha}
(\nu_L)_j](\widetilde{d}_L)_{k\alpha}\right\} \ , 
\eqno(2.1)
$$
where $\psi^c\equiv C \overline{\psi}^T$ and the indices
$(i,j,\cdots)$, $(A,B,\cdots)$ and $(\alpha,\beta,\cdots)$
are family-, SU(5)$_{GUT}$- and SU(3)$_{color}$-indices,
respectively.
However, in GUT models, if the interactions (2.1) exist, 
the following $R$-parity violating interactions will also 
exist:
$$
\lambda_{ij}^k(\overline{\psi}_{\overline{5}}^c)_i^A
(\psi_{10})_{kAB}
(\widetilde{\psi}_{\overline{5}})_{j}^B
$$
$$
=\frac{1}{\sqrt{2}}\lambda_{ij}^{k}
\left\{ \varepsilon_{\alpha \beta \gamma}
(\overline{d}_R)_i^{\alpha}
(\widetilde{d}_R^{\dagger})_{j}^{\beta}
(u_R^{c})_k^{\gamma} \right.
$$
$$
-[(\overline{e}_L^c)_i(\widetilde{\nu}_L)_j 
- (\overline{\nu}_L^c)_i
(\widetilde{e}_L)_j](e_R^c)_k
$$
$$
- [(\overline{e}_L^c)_i(\widetilde{d}_
R^{\dagger})_j^{\alpha}
- (\overline{d}_R)_i^\alpha
(\widetilde{e}_L)_j] (u_L)_{k\alpha}
$$
$$ 
\left.
+ [(\overline{\nu}_L^c)_i
(\widetilde{d}_R^{\dagger})_j^{\alpha} -
(\overline{d}_R)_i^{\alpha}
(\widetilde{\nu}_L)_j](d_L)_{k\alpha}
\right\} \ ,
\eqno(2.2)
$$
which contribute to the proton
decay through the intermediate state
$\widetilde{d}_R$. 

In order to forbid the contribution
of the interactions (2.2)
to the proton decay, for example, we can assume that 
the $R$-parity violating interactions occur
only when the field $\psi_{10}$ of the
third family is related,
i.e., we assume the interactions 
$$
\lambda_{ij}^3 (\overline{\psi}_{\overline{5}}^c)_i^{A}
(\psi_{10})_{3AB}(\widetilde{\psi}_{\overline{5}})_j^B \ ,
\eqno(2.3)
$$
instead of the interaction (2.2).
Then, the terms
$\lambda_{ij}^3(\overline{d}_R)_i
(\widetilde{d}_R^{\dagger})_j
(u_R^c)_3$ cannot contribute to the proton decay.
In order to realize the constraints 
$$
\lambda^k_{12}=\lambda^k_{23}=\lambda^k_{31}=0  \ \ \  
{\rm for}\ \  k=1,2 \ ,
\eqno(2.4)
$$
we introduce a 
discrete symmetry Z$_2$,
which exactly holds at every energy scale, as follows:
$$
\begin{array}{ll}
(\psi_{\overline{5}})_i \rightarrow
\eta_i (\psi_{\overline{5}})_i \ , &
(\widetilde{\psi}_{\overline{5}})_i \rightarrow 
 \eta_i (\widetilde{\psi}_{\overline{5}})_i \ , \\
(\psi_{10})_i \rightarrow  \xi_i (\psi_{10})_i \ , &
(\widetilde{\psi}_{10})_i \rightarrow 
 \xi_i (\widetilde{\psi}_{10})_i \ , \\
\end{array}
\eqno(2.5)
$$
where $\eta_i$ and $\xi_i$ take 
$$
 \eta=(+1,+1,+1) \ , \ \ \xi=(-1,-1,+1) \ , 
\eqno(2.6)
$$
under the Z$_2$ symmetry.
Then, the Z$_2$ invariance leads to
the constraints (2.4).

However, if the RGE effects cause a mixing between the 
first and third families,
the interactions (2.3) can again
contribute to the proton decay. 
If we assume that ${\bf 5}$ and
$\overline{{\bf 5}}$ Higgs fields
$H_u$ and $H_d$ transform as
$$
H_u \rightarrow + H_u \ , \ \ \ H_d \rightarrow +H_d \ ,
\eqno(2.7)
$$
under the Z$_2$ symmetry, the up-quark mass matrix $M_u$
is given by the form
$$
M_u = \left(
\begin{array}{ccc}
c_u & d_u & 0 \\
d_u & b_u & 0 \\
0 & 0 & a_u 
\end{array} \right) \ .
\eqno(2.8)
$$
This guarantees that the top quark $u_3$ in the
$R$-parity violating terms (2.3) does not mix with the
other components ($u_1$ and $u_2$) even if we take
the RGE effects into consideration, so that the 
interactions (2.3) cannot contribute to the proton
decay at any energy scales.

On the other hand, the down-quark mass matrix $M_d$
and the charged lepton mass matrix $M_e$, which are
generated by the Higgs scalar $H_d$, have the form
$$
M_d = M_e^T =\left(
\begin{array}{ccc}
0 & 0 & 0 \\
0 & 0 & 0 \\
a_1 & a_2 & a_3 
\end{array} \right) \ .
\eqno(2.9)
$$
The mass matrix form (2.9) cannot explain the observed
masses and mixings.
In order to give reasonable masses and mixings of the quarks
and charged leptons, we must consider additional SU(5) 
45-plet Higgs scalars, which do not contribute to the
up-quark mass matrix because 
$\overline{\psi}_{10} M_u\psi_{10}^c$ belongs
to $(\overline{10}\times \overline{10})_{symmetric}$.
Then, we obtain the down-fermion mass matrices
$$
M_d = \left(
\begin{array}{ccc}
c'_1 & c'_2 & c'_3 \\
b'_1 & b'_2 & b'_3 \\
a_1+a'_1 & a_2+a'_2 & a_3+a'_3 
\end{array} \right) \ , 
\eqno(2.10)
$$
$$ 
M_e^T= \left(
\begin{array}{ccc}
-3c'_1 & -3c'_2 & -3c'_3 \\
-3b'_1 & -3b'_2 & -3b'_3 \\
a_1-3a'_1 & a_2-3 a'_2 & a_3-3a'_3 
\end{array} \right) \ ,
\eqno(2.11)
$$
where $a'_i$ and $(b'_i, c'_i)$ denote contributions
from the 45-plet Higgs scalars $H_{45}^{(+)}$ and
$H_{45}^{(-)}$ which transform $H_{45}^{(+)} \rightarrow
+H_{45}^{(+)}$ and $H_{45}^{(-)} \rightarrow
-H_{45}^{(-)}$ under the symmetry Z$_2$, respectively.
Note that if we consider either $H_{45}^{(+)}$ or
$H_{45}^{(-)}$, we cannot still give realistic 
down-fermion masses.  
For example, if we have $H_5$ ($H_u$ and $H_d$) and
$H_{45}^{(-)}$, we obtain an unwelcome relation
$$
m_\tau^2 +m_\mu^2+m_e^2 -(m_b^2+m_s^2+m_d^2)
$$
$$
=8 \sum \left( |b'_i|^2 +|c'_i|^2\right) >0 \ ,
\eqno(2.12)
$$
from the trace of $M_e^\dagger M_e - M_d M_d^\dagger$.
Therefore, in order to obtain realistic down-fermion
mass matrices, we need, at least, the three types of
the Higgs scalars $H_5$, $H_{45}^{(+)}$
and $H_{45}^{(-)}$.

However, such additional Higgs scalars $H_{45}^{(\pm)}$
cause another problems.
One is a problem of the flavor changing neutral currents
(FCNC).  This problem is a common subject to overcome 
not only in the present model but also in most GUT models.  
The conventional mass matrix models 
based on GUT scenario cannot give realistic mass 
matrices without assuming Higgs scalars more than two.
For this problem, we optimistically consider that only
one component of the linear combinations among those
Higgs scalars survives at the low energy scale 
$\mu=\Lambda_L$ ($\Lambda_L$ is the electroweak energy
scale), while other components are decoupled at
$\mu < \Lambda_X$ ($\Lambda_X$ is a unification scale).

Another problem is that the {\bf 45} Higgs
scalars can have  vacuums expectation values (VEV) 
at the electroweak energy scale $\Lambda_L$, 
so that the Z$_2$ symmetry is broken 
at $\mu = \Lambda_L$.
Therefore, the proton decay may occur through higher
order Feynman diagrams.  
In the conventional GUT models, it is still a current
topic whether the colored components of the SU(5)
5-plet Higgs scalar can become sufficiently heavy or
not  to suppress the proton decay. 
We again optimistically assume that the colored 
components of the 45-plet Higgs scalars are sufficiently
heavy to suppress the proton decay, i.e.,
that such effects will be suppressed by a factor 
$(\Lambda_L/\Lambda_X)^2$.

\section{Radiatively induced neutrino masses}

We define fields $u_i$, $d_i$ and $e_i$ as
those corresponding to mass
eigenstates, i.e.,
$$
H_{mass} = \overline{u}_L U_L^{u \dagger} M_u U_R^u u_R 
+ \overline{d}_L U_L^{d \dagger} M_d U_R^d d_R 
$$
$$
+  \overline{e}_L U_L^{e \dagger} M_e U_R^e e_R + h.c. \ ,
\eqno(3.1)
$$
and fields $\nu_{Li}$ as partners of 
the mass eigenstates $e_{Li}$, i.e., 
$\ell_{Li}=(\nu_{Li} , e_{Li})$.
We define the neutrino mass matrix $ M_{\nu} $ as 
$$
H_{\nu\ mass} = \overline{\nu}_L^c M_\nu \nu_L \ .
\eqno(3.2)
$$
Therefore, a unitary matrix $U_L^\nu$ which is defined by
$$
U_L^{\nu T} M_\nu U_L^\nu = D_\nu \equiv
 {\rm diag}(m_1^\nu, m_2^\nu, m_3^\nu) \ ,
\eqno(3.3)
$$
is identified as the
Maki-Nakagawa-Sakata-Pontecorvo \cite{MNS}
neutrino mixing matrix 
$U_{MNSP}=U_L^\nu$.


\begin{figure}
\includegraphics[width=65mm]{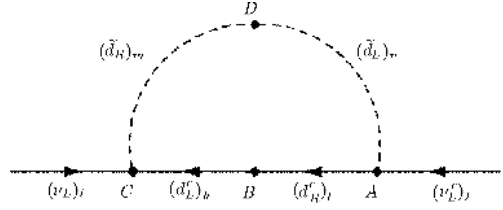} 
\caption{
Radiatively induced neutrino mass through the
down-quark loop. The vertices A, B, C and D are
given by
$(U_R^{d\dagger} \lambda U_L^{e})_{lj} 
(\widetilde{U}_L^d)_{3n}$,
$(U_L^{d\dagger} M_d U_R^d)_{kl}$, 
$(\widetilde{U}_R^{d\dagger} \lambda U_L^e)_{mi} 
({U}_L^d)_{3k}$, and $(\widetilde{U}_L^{d\dagger} 
\widetilde{m}_d^2 \widetilde{U}_R^d)_{kl}$,
respectively.
}
\label{radm-d}
\end{figure}

In addition to the $R$-parity
violating terms (2.1) and (2.2) [(2.3)],
we assume SUSY breaking terms
$\widetilde{\psi}_{\overline{5}}
\widetilde{\psi}_{10} H_{\overline{5}}^d$ (and 
$\widetilde{\psi}_{\overline{5}}
\widetilde{\psi}_{10} H_{\overline{45}}^d$).
For the moment, we do  not consider
$\widetilde{e}_R^c$-$H_d^+ $ mixing
as in the original Zee model. 
(The $H_d^+$ contribution to the neutrino mass matrix
will be discussed in the end of this section.)
Then, the neutrino masses are radiatively generated.
In Fig.~\ref{radm-d}, we illustrate the
 Feynman diagram for the case
with the down-quark loop.
The amplitude is proportional  to the coefficient 
$$
(U_R^{d \dagger} \lambda U_L^{e})_{lj}
(\widetilde{U}_L^d)_{3n}
\cdot (U_L^{d \dagger} M_d U_R^d)_{kl} 
$$
$$
\cdot
(\widetilde{U}_R^{d \dagger} \lambda^T U_L^e)_{mi}
({U}_L^d)_{3k}
\cdot (\widetilde{U}_L^{d \dagger} \widetilde{m}_d^2 
\widetilde{U}_R^d)_{nm} 
$$
$$
=(\widetilde{m}_d^2 \lambda U_L^e)_{3i}
(M_d \lambda U_L^{e})_{3j} \ ,
\eqno(3.4)
$$
where $(\widetilde{m}^2_d)_{ij}$ are
coefficients of $(\widetilde{d}_L^\dagger)_i
(\widetilde{d}_R)_j$,
and $(\lambda)_{ij}=\lambda^3_{ij}$.
Similarly, we obtain the contributions from
the charged lepton loops.
Therefore, the radiatively induced neutrino
mass matrix $M_\nu$ is given by
the following form
$$
(M_\nu)_{ij} = (f_i^e g_j^e+f_j^e g_i^e) K_e +
(f_i^d g_j^d+f_j^d g_i^d) K_d \ ,
\eqno(3.5)
$$
where $K_f$ ($f=e,d$) are common factors independently
of the
families, and
$$
\begin{array}{l}
f_i^e = (M_e^T \lambda U_L^{e})_{3i} \ , \ \ \ 
g_i^e = (\widetilde{m}_e^{2T} \lambda U_L^e)_{3i} \ , \\
f_i^d = (M_d \lambda U_L^{e})_{3i} \ , \ \ \ 
g_i^d = (\widetilde{m}_d^2 \lambda U_L^e)_{3i} \ . \\
\end{array}
\eqno(3.6)
$$

On the other hand, the contributions due to the 
$H_d^+$ - $\widetilde{e}^+$ mixing are as follows.
There are no contributions from bilinear terms
$H_u(5)[H_d (\overline5) + \sum_{i}\widetilde{\psi}(\overline5)]$,
because we can always eliminate the contributions from 
$H_u(5)\widetilde{\psi}(\overline5)$ 
by re-definition of the scalar
$H_d(\overline5)$. Also, there are no contributions from
$H_d (\overline5)H_d(\overline5)\widetilde{\psi}(10)$,
because $\widetilde{\psi}(10)$ is anti-symmetric in SU(5) indices.
However, the terms 
$H_d (\overline5)H_d(\overline{45})\widetilde{\psi}(10)$ 
can cause the $H_d^+$ - $\widetilde e ^+$
mixing. These contributions affect only to the 
charged lepton loop diagrams.

The final result which includes the $H_d ^+$
contributions is given by
$$
M_{ij}=
(f_i^e g_j^e + f_j^e g_i^e) K_e
+ (f_i^d g_j^d + f_j^d g_i^e) K_d
+F_{ij}^e K_e^{'} ,
\eqno(3.7)
$$
where
$$
F_{ij}^e=
({U_L^e}^T \lambda M_e {M_e^{'}}^\dagger U_L^e)_{ij}
+(i \leftrightarrow j) ,
\eqno(3.8)
$$
and $M_e^{'}$ denotes the contributions due to 
$\sum \overline\nu_L e_L H_d^+ $
[$H_d^+$ denotes $(H_d (\overline5))^+$,
$(H_d^{(+)} (\overline{45}))^+$
and $(H_d^{(-)} (\overline{45}))^+ $].
We have already assumed that only one component of the
linear combinations among those Higgs scalars survives 
at the low energy scale.
Then, we can regard $M_e^{'}$ as $M_e^{'}=M_e$, so that 
we obtain $F_{ij}^e$ as follows:
$$
F_{ij}^e=
(\lambda^{'} D_e^2)_{ij} +(i \leftrightarrow j)
=\lambda_{ij}^{'} ({m_j^e}^2 - {m_i^e}^2) ,
\eqno(3.9)
$$
where $\lambda^{'}={U_L^e}^T \lambda U_L^e $ is
antisymmetric tensor as well as $\lambda$ ,
and $m_i^e = (m_e , m_\mu , m_\tau).$
Note that, in this case, the third term in (3.7)
has the same mass matrix form as that in the original
Zee model.

\section{Phenomenology}

In the SUSY GUT scenario, there are many origins of the 
neutrino mass generations.
For example, the sneutrinos $\widetilde{\nu}_{iL}$ can
have the VEV, and thereby, the neutrinos $\nu_{Li}$ acquire
their masses (for example, see Ref.\cite{Diaz}).
Although we cannot rule out a possibility that the observed
neutrino masses can be understood from such compound origins,
in the present paper, we do not take such the point of view, 
because the observed neutrino masses and mixings appear to be 
rather simple and characteristic.
We simply assume that 
the radiative masses are only dominated even if there are
other origins of the neutrino mass generations.

However, even under such the assumption, we still have many 
parameters as shown in the expression (3.7).
For simplicity, we assume that the contributions from the
charged lepton loop are dominated compared with those from
the down-quark loop, i.e., $K_e \gg K_d$, which corresponds
to the case ${m}^2(\widetilde{d}_L) - {m}^2(\widetilde{d}_R)
\gg {m}^2(\widetilde{e}_L) -{m}^2(\widetilde{e}_R)$.
(We can give a similar discussion for the case $K_d \gg K_e$.
In the present paper, we do not discuss which case is
reasonable.)
Furthermore, we neglect the third term ($H_d^+$ contributions)
in (3.7).
Then, the neutrino mass matrix
(3.5) becomes a simple form 
$$
(M_{\nu})_{ij} = m_0 (f_ig_j + f_jg_i).
\eqno(4.1)
$$
Hereafter, for convenience, we will normalize $f_i$
and $g_i$ as
$$
|f_1|^2 + |f_2|^2 + |f_3|^2 = 1,
\ \ |g_1|^2 + |g_2|^2 + |g_3|^2 = 1 \ .
\eqno(4.2)
$$
In the most SUSY models, it is taken that the form
of $ \widetilde {m}^2_f$ $(f=e, d)$ is proportional to the
fermion mass
matrix $M_f$. 
Then, the coefficients $ g_i$ are proportional to $f_i$,
so that the mass matrix (4.1) becomes $ (M_{\nu})_{ij} =
2 m_0 f_if_j$,
which is a rank one matrix. 
Therefore, we rule out the case with $\widetilde{m}_f^2
\propto M_f$.

For convenience, hereafter, we assume that $f_i$ and $g_i$ 
($i=1,2,3$) are real. 
The mass eigenvalues and mixing matrix elements for the neutrino 
mass matrix (4.1) are given as follows :
$$
\begin{array}{l}
m_1^\nu=(1+\varepsilon)m_0 \ , \\
m_2^\nu=-(1-\varepsilon)m_0 \ ,  \\
m_3^\nu=0 \ ,
\end{array} 
\eqno(4.3)
$$
$$
\renewcommand{\arraystretch}{2}
\begin{array}{l}
U_{i1}=\displaystyle\frac{1}{\sqrt{2}}
\displaystyle\frac{f_i+g_i}{\sqrt{1+\varepsilon}} \ , \\
U_{i2}=\displaystyle\frac{1}{\sqrt{2}}
\displaystyle\frac{f_i-g_i}{\sqrt{1-\varepsilon}} \ , \\
U_{i3}=- \varepsilon_{ijk}
\displaystyle\frac{f_jg_k}{\sqrt{1-\varepsilon^2}} \ , \\
\end{array}
\eqno(4.4)
\renewcommand{\arraystretch}{1}
$$
where
$$
\varepsilon= f_1g_1 + f_2g_2 + f_3g_3 \ .
\eqno(4.5)
$$

As seen in (4.3), the mass level pattern of the present 
model shows the inverse hierarchy as well as that of the 
Zee model. 
{}From (4.3), we obtain
$$
\begin{array}{l}
\Delta m_{21}^2 \equiv (m_2^\nu)^2 - (m_1^\nu)^2 = 
- 4\varepsilon m_0^2 \ , \\
\Delta m_{32}^2 \equiv (m_3^\nu)^2 - (m_2^\nu)^2 = 
- (1-\varepsilon)^2 m_0^2 \ , \\
\end{array} 
\eqno(4.6)
$$
$$
R \equiv \frac{\Delta m_{21}^2}{\Delta m_{32}^2}=
\frac{4\varepsilon}{(1-\varepsilon)^2} \ .
\eqno(4.7)
$$

 For a small $R$, the mixing parameters $\sin^2 2\theta_{solar}$, 
$\sin^2 2\theta_{atm}$ and $U_{e3}^2$ are given by
$$
\sin^2 2 \theta_{solar} \equiv 4U_{11}^2U_{12}^2=
\frac{1}{1-\varepsilon^2}
(f_1^2-g_1^2)^2 \ , 
\eqno(4.8)
$$
$$
\sin^2 2\theta_{atm} \equiv 4 U_{23}^2 U_{33}^2 
$$
$$
= \frac{4}{(1-\varepsilon^2)^2}
[f_2f_3 + g_2g_3 - \varepsilon(f_3g_2 + f_2g_3)]^2 \ ,
\eqno(4.9)
$$
$$
U_{13}^2       
= 1 - \frac{f_1^2 + g_1^2 - 2\varepsilon f_1g_1}{1-\varepsilon^2}
 \ .
\eqno(4.10)
$$

Let us demonstrate that the mass matrix form (4.1)
has  reasonable
parameters for the observed neutrino data.
The atmospheric neutrino data \cite{atm} suggests a large 
$\nu_\mu$-$\nu_\tau$ mixing, i.e., $\sin^2 2\theta_{23}
\simeq 1$.
It is known that the nearly bimaximal mixing is derived from
the neutrino mass matrix $M_\nu$ with $2\leftrightarrow
3$ symmetry
\cite{Fukuyama}. 
Therefore, we take 
$$
\begin{array}{ll}
f_1 = s_\alpha , & f_2 = f_3 = \frac{1}{\sqrt{2}}
c_\alpha \ ,\\
g_1 = c_\beta , & g_2 = g_3 = -\frac{1}{\sqrt{2}}
s_\beta \ ,\\
\end{array}
\eqno(4.11)
$$
where $c_\alpha= \cos\alpha$, $s_\alpha=\sin\alpha$
and so on.
Then, the parameterization (4.11) gives
$$
\varepsilon = \sin(\alpha -\beta)\ ,
\eqno(4.12)
$$
$$
\sin^2 2\theta_{solar} = \cos^2(\alpha +\beta)
\ , \eqno(4.13)
$$
$$
\sin^2 2\theta_{atm} = 1
\ , \eqno(4.14)
$$
$$
U_{13}^2 = 0 \ .
\eqno(4.15)
$$
We assume that the values of $\alpha$ and $\beta$ are
highly
close each
other, i.e., $\sin(\alpha-\beta) \sim 10^{-2}$.
The result (4.13) with $\alpha\simeq \beta$ is free from
the constraint
(1.1) in the original Zee model, so that we can fit the
value of
$\sin^2 2\theta_{solar}$
with the observed value \cite{solar}
$\sin^2 2\theta_{solar} \sim 0.8$ from the solar neutrino data 
by adjusting
the parameter $\alpha$ ($\simeq \beta$).
Of course, the parameterization (4.11) is taken only from the
phenomenological point of view, so that the
results (4.13)-(4.15) are
not theoretical consequences in the present model.

{}From the recent atmospheric and solar neutrino data
\cite{atm,solar} $R \simeq (4.5 \times 10^{-5}$ eV$^2)
/(2.5 \times 10^{-3})$ eV$^2$)$= 1.8 \times 10^{-2}$,
we estimate $\varepsilon =4.5 \times 10^{-3}$, and
$m_0 \simeq m_1^\nu \simeq |m_2^\nu| \simeq \sqrt
{\Delta m_{atm}^2 }
= 0.050 \ {\rm eV}$.
The effective neutrino mass $\langle m_\nu \rangle$ from the
neutrinoless double beta decay experiment is given by
$\langle m_\nu \rangle =(M_\nu )_{11} =
2 m_0 c_\alpha s_\beta 
\simeq m_0 \sqrt{ 1-\sin^2 2\theta_{solar} } 
\simeq 2.2 \times 10^{-3} \ {\rm eV}$,
where we have used the observed value \cite{solar}
$\sin^2 2\theta_{solar} \simeq 0.8$.

\section{Conclusion}

In conclusion, we have proposed a neutrino
 mass matrix model
based on a SUSY GUT model where only top quark takes
$R$-parity violating interactions and the
Z$_2$ symmetry
plays an essential role, so that the proton decay
due to the $R$-parity interactions can be
evaded safely.

The general form of the radiatively induced neutrino
mass matrix in the present model is given by the
expression (3.7). Since the form (3.7) has too many
parameters, we have investigated phenomenology for a
simplified case (4.1). Although we have investigated
the case $\varepsilon \simeq 0$ [$\varepsilon$ is
defined by (4.5)], the case $1- \varepsilon \simeq 0$
is also possible.  However,
the purpose of the present paper is to give the general
form of the radiatively induced neutrino mass matrix
(3.7) under the discrete symmetry Z$_2$.
A systematic study of the mass matrix (3.7) will be
given elsewhere.

\vspace*{3mm}
{\bf Acknowledgements}

The authors would like to thank to M.~Yasue
and J.~Sato
for their helpful discussions and valuable
comments.
They also thank to J.~W.~F.~Valle and
 D.~Marfatia for
their useful comments and informing
valuable references.
One of the authors (AG) is supported
by the Japan Society
for Promotion of Science (JSPS)
Postdoctoral Fellowship
for Foreign Researchers in Japan
through Grant No.P99222.



\end{document}